\documentclass[11pt,a4paper,titlepage,openbib]{article}
\usepackage{caption}
\usepackage{subfig}
\usepackage{amsmath}
\usepackage{amsfonts}
\usepackage{amssymb}
\usepackage[pdftex]{graphicx}

\author{Deepak Chandran}
\begin{document}

\begin{titlepage}
\begin{center}
\ \\ 
\textsc{\Large
Bioengineering Department, \\
University of Washington, Seattle, WA\\
\ \\
\ \\
\ \\
\ \\
\ \\}
%
%
{\huge \bfseries
An Optimization Algorithm for Finding Parameters for Bistability
\\}
\ \\
\ \\
\ \\
\ \\
\begin{minipage}{0.4\textwidth}
\begin{flushleft} \large
\emph{Authors:}\\
Deepak Chandran$^1$ \\
Herbert M. Sauro$^1$  \\
\ \\
\ \\
\small
$^{1}$Department of Bioengineering, University of Washington, Box 355061, William H. Foege Building, Room N210E, Seattle, WA, USA 98195-5061 \\
\end{flushleft}
\end{minipage}
%
\begin{minipage}{0.4\textwidth}
\begin{flushright} \large
\footnotesize{\ }
\end{flushright}
\end{minipage}
\vfill 
%
{\large \today}
\end{center}
\end{titlepage}



\section*{Abstract}

\textbf{Motivation: }

Many biochemical pathways are known, but the numerous parameters required to correctly explore the dynamics of the pathways are not known. For this reason, algorithms that can make inferences by looking at the topology of a network are desirable. In this work, we are particular interested in the question of whether a given pathway can potentially harbor multiple stable steady states. In other words, the challenge is to find the set of parameters such that the dynamical system defined by a set of ordinary differential equations will contain multiple stable steady states. Being able to find parameters that cause a network to be bistable may also be benefitial for engineering synthetic bistable systems where the engineer needs to know a working set of parameters.

\textbf{Result: }

We have developed an algorithm that optimizes the parameters of a dynamical system so that the system will contain at least one saddle or unstable point. The algorithm then looks at trajectories around this saddle or unstable point to see whether the different trajectories converge to different stable points. The algorithm returns the parameters that causes the system to exhibit multiple stable points. Since this is an optimization algorithm, it is not quaranteed to find a solution. Repeated runs are often required to find a solution for systems where only a narrow set of parameters exhibit bistability.

\textbf{Availability: }

The C code for the algorithm is available at http://tinkercell.googlecode.com.

\section{Introduction}

The existence of more than one stable state is a common theme in biochemical networks. Bistable networks, or networks with two stable steady states, have been observed to be very common in biology and may have great importance in the overall behavior of a single cell or multi-cell system. Theoretical methods have been developed to determine the possibility of bistability by looking at network topology \cite{feinberg1972ckc, feinberg1972cbg, feinberg1987crn, feinberg1995eau, craciun2005mec, craciun2006ubc}. However, these methods are usually limited to networks that obey mass-action kinetics, but a vast majority of models use more complex kinetics. Other theoretical methods also exist that use certain criteria to guarantee the presence of multiple stable states \cite{angeli2004msm}. These methods are based on the understanding that a positive feedback is a neccessary condition for bistability \cite{thomas1990bf,thomas1995dbb}, and monotonicity, in addition to positive feedback, can be a sufficient condition. However, such methods can be stringent, thus missing some of the bistable systems. A numerical method has also been developed that uses the eigenvalues of a system to search for bifurcation points, thus identifying the point of transition from monostability to bistability \cite{chickarmane2005bdt}. The algorithm presented in this work is also a numerical method, but it uses a slightly different method to optimize the system which shows improvement over the aforementioed algorithm.

\section{Method}
The goal of the algorithm presented in this work is as follows:

\begin{flushleft}
\textbf{Given:} A system of differential equations, $d\overline{X}/dt = F(\overline{X},P)$ with an unknown parameter set $P = { k_{1}...k_{n} }$

\textbf{Find:} $P$ such that $d\overline{X}/dt = F(\overline{X},P)$ has more than one stable steady state. We define a stable steady state as the point where a time-course simulation converges and where all the eigenvalues of the Jacobian matrix are negative. 
\end{flushleft}

We postulate that a system with more than one stable point will have at least one unstable or saddle point. The argument for this statement can be understood if one considers the energy landscape of a dynamical system. The energy landscape is a hypothetical phase plane where the ``energy" at each point is proportional to the probability of the system existing at that point at any given time. The valleys, or low energy regions, of such an energy landscape correspond to the stable points, and the peaks, or high energy regions, correspond to the unstable or saddle points. One can argue that if there are two valleys, there will exist a peak between them. Therefore, a system with two or more stable points will have an unstable or saddle point between them. However, one should note that the existence of a peak does not acertain existence of two valleys. However, for the algorithm presented in this work, we assume that the majority of the cases for biological systems will be such that the existence of a peak also implies the existence of two or more valleys. Under this assumption, one can find bistable systems by finding systems with one unstable or saddle point. This is the goal of the algorithm presented in this work: search for a system that contains an unstable of saddle point. Once this point, or peak, is found, the two stable points, or valleys, can be found by following the trajectories around the unstable or saddle point. 

Given the above general idea, one can restate the goal of the algorithm differently: find $P$ such that $d\overline{X}/dt = F(\overline{X},P)$ has at least one unstable or saddle point. 

Once the unstable or saddle point is found, the existence of more than one stable point can be confirmed through time-course simulations around the unstable or saddle point. In order to find the parameters that cause the system to harbor at least one saddle or unstable point, the algorithm optimizes the following objective function:

\begin{flushleft}
\textbf{Given:} A system of differential equations, $d\overline{X}/dt = \alpha F(\overline{X},P)$ with an unknown parameter set $P = { k_{1}...k_{n} }$ and vector $\alpha$. $\alpha$ is a vector of reals and has the same size as $\overline{X}$.

\textbf{Find:} $P$ and $\alpha$ such that not all values in $\alpha$ are positive and $d\overline{X}/dt = \alpha F(\overline{X},P)$ has at least one stable steady state.
\end{flushleft}

Since the $\alpha$ vector described above is not all positive, the \textit{stable} steady state of the system $d\overline{X}/dt = \alpha F(\overline{X},P)$ will be a \textit{saddle or unstable} steady state for the original system $d\overline{X}/dt = F(\overline{X},P)$. Note that the steady state solutions for the system with or without $\alpha$ is the same since $\alpha$ is a vector of constants. However, since at least one of the $\alpha$ values is negative, the stability of each steady state will be different for the system with and without $\alpha$. For a monostable system, the system with $\alpha$ will not converge at all, because the only steady state in the system will be a saddle or unstable point. For a bistable system, there will be some $\alpha$ such that the system with $\alpha$ will converge to the saddle or unstable point of the original system. Therefore the parameter set, $P$, that causes $d\overline{X}/dt = \alpha F(\overline{X},P)$ to converge will cause $d\overline{X}/dt = F(\overline{X},P)$ to be bistable.

\subsection{Numerical libraries used}

We used a genetic algorithm to optimize the objective function. The genetic algorithm C library was obtained from http://tinkercell.googlecode.com. The CVODE library in the Sundials package was used to perform numerical integration. The optimization was performed using a Nelder-Mead C library obtained from \\ http://www.ritsumei.ac.jp/se/~hirai/edu/2003/algorithm/index-e.html. \\ The eigenvalue calculations were performed using CLAPACK. 

\section{Results}

We tested the algorithm using a few known bistable systems. One of the systems that was tested was:

$x' =  \alpha_{1} (k_2 - k_3 x + k_0 x^2y - k_1 x^3)$\\
$y ' = \alpha_{2} (k_4 - k_5 y - (k_0 x^2y - k_1 x^3))$

The optimization returned the following values:
where the $\alpha = [-0.3995 , 0.917]$ and $k = [4.27 , 2.39 , 0.007 , 5.55 , 4.44 , 0.5]$

\begin{figure}
  \vspace{-10pt}
   \centering
    \subfloat[unstable steady state]{\label{fig:phase1a}\includegraphics[width=0.5\textwidth]{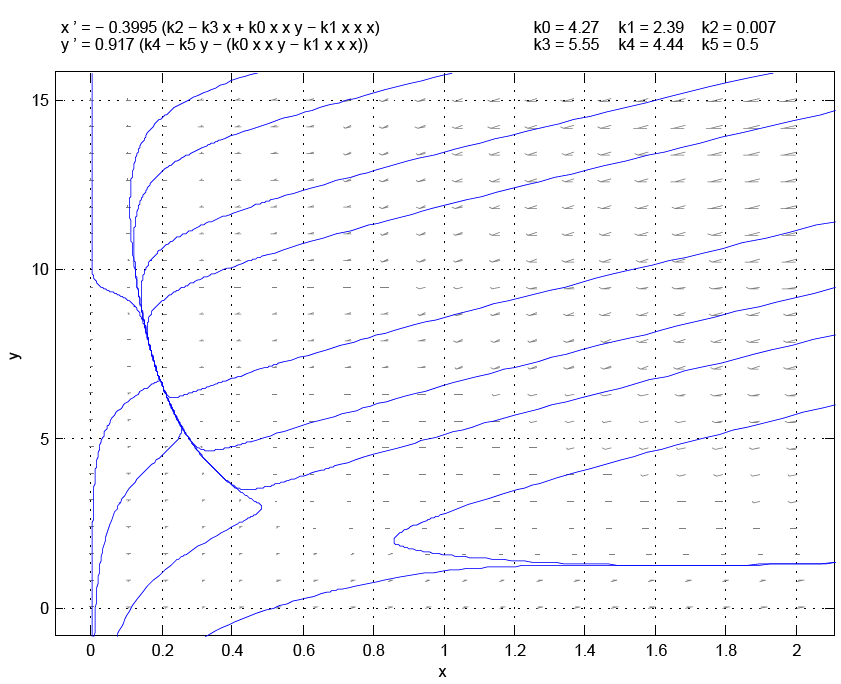}}                
    \subfloat[stable steady states]{\label{fig:phase1b}\includegraphics[width=0.5\textwidth]{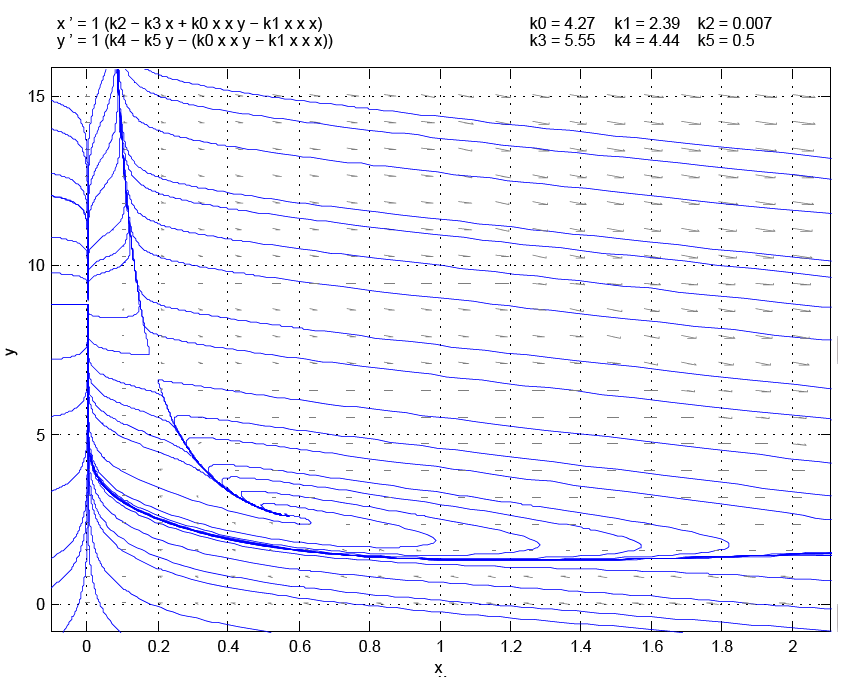}}
  \vspace{-10pt}
  \caption{ (a) The phase portrait of a 2-variable system when one of the $\alpha$ values is negative. (b) The phase portrait of the same variable when all of the $\alpha$ values are 1. Note that the stable point in (a) is the unstable point in (b). }\label{fig:phase1}
\end{figure}

The system above is shown in Figure \ref{fig:phase1a} and \ref{fig:phase1b}. The $\alpha$ and $k$ stated above were found by forcing the system to have one stable state. Note that $\alpha$ has at least one negative value. The stable point in Figure \ref{fig:phase1a} is the saddle point in \ref{fig:phase1b}, which will always be the case. The phase portrait in Figure \ref{fig:phase1b} is the same system as Figure \ref{fig:phase1a} except with $\alpha = [1, 1]$, i.e. the normal system. The normal system with the parameter set $k$ is bistable because the system with $\alpha$ has a stable point. 

Another system that was tested is a genetic toggle switch \cite{gardner2000cgt}:

$x' = \alpha_1 (k0/(k_1 + y^4) - k_2 x)$\\
$y ' = \alpha_2 (k3/(k_4 + x^4) - k_5 y)$

The optimization returned the following values:
where the $\alpha = [-1.0 , 2.5]$ and $k = [2.8, 0.5 , 1.04, 2.2 , 0.25 , 1.05]$

\begin{figure}
  \vspace{-10pt}
   \centering
    \subfloat[unstable steady state]{\label{fig:phase2a}\includegraphics[width=0.5\textwidth]{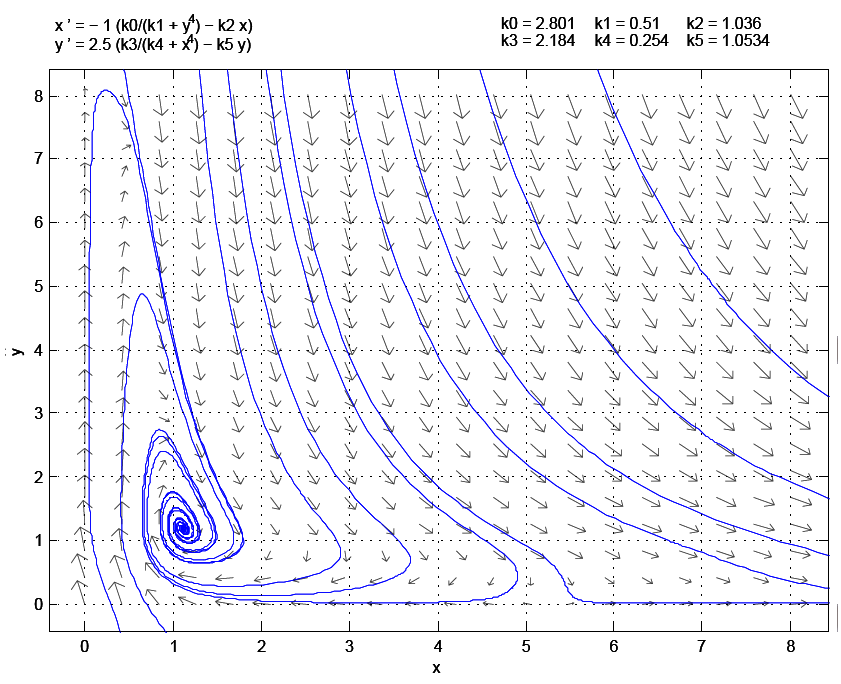}}                
    \subfloat[stable steady state]{\label{fig:phase2b}\includegraphics[width=0.5\textwidth]{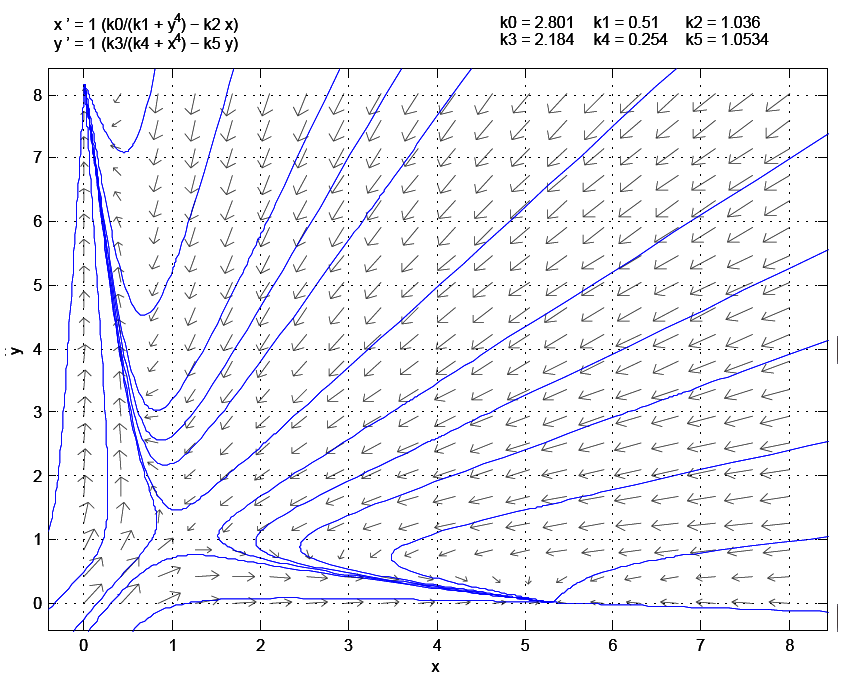}}
  \vspace{-10pt}
  \caption{(a) The phase portrait of a 2-variable system when one of the $\alpha$ values is negative. (b) The phase portrait of the same variable when all of the $\alpha$ values are 1. Note that the stable point in (a) is the unstable point in (b).}\label{fig:phase2}
\end{figure}

Again, since the system with $\alpha$ has a stable point, the system without $\alpha$ has a saddle point and two stable points. The phase portraits are shown in Figure \ref{fig:phase2}.

The algorithm is not always successful in finding the correct parameters. For the systems described above, the algorithm is generally successful once in every two runs. It takes roughly 20 iterations to find the parameters, which is an average of 20 seconds (1 second per iteration) on a computer using AMD Opteron 165 1.8GHz. 

Previous methods were unsuccessful to identify bistability in both of the systems. The second system (Figure \ref{fig:phase2}) does not use mass-action kinetics, so methods that rely on mass-action kinetics will fail. The first system (Figure \ref{fig:phase1}) is not always monotonic, so using monotonicity and positive feedback as a predictor for bistability will fail. Previous numeric methods (ref) were also unsuccessful in finding correct parameters for either of the systems.

\section{Conclusions}

We have presented an algorithm that can find parameters for a system such that the system will contain more than one stable steady state. The application for such an algorithm would be for systems, particularly biological systems, where parameters are generally not known. Using this algorithm, one can ``guess" that the system can potentially be bistable. If the bistability of the system holds biological meaning, then it is possible that the system is indeed bistable. 

The algorithm presented in this work is not guaranteed to find the parameters for bistability, which can be an issue for systems where the range of parameters for bistability is small. However, we hope that the general concept from this algorithm can be modified and improved. For example, the genetic algorithm that is used for optimization can be replaced with a different algorithm. The objective function can be restated to improve the search process. Such changes may improve the success rate of the algorithm. 

\section{Acknowledgements}
This work was funded by the National Science Foundation (Id 0527023- FIBR).

\pagebreak 
\section{Literature Cited}
\renewcommand{\refname}{}
\bibliographystyle{bmc_article}
\bibliography{ref}
\pagebreak 

\end{document}